\documentclass[preprint2]{aastex}
\usepackage{graphicx}

\begin{document}

\title{Application of algorithms for high precision metrology}

\author{M. Gai\altaffilmark{1}, A. Riva\altaffilmark{1}, 
D. Busonero\altaffilmark{1}, R. Buzzi\altaffilmark{1}, 
F. Russo\altaffilmark{1}}

\affil{ Istituto Nazionale di Astrofisica - Osservatorio Astrofisico 
di Torino, \\ 
V. Osservatorio 20, 10025 Pino T.se (TO), Italy }

\email{gai@oato.inaf.it}

\shorttitle{Application of algorithms for high precision metrology}
\shortauthors{Gai et al.}

\begin{abstract}
This paper evaluates the performance of algorithms suitable to process 
the measurements from two laser beam metrology systems, in particular 
with reference to the Gaia Basic Angle Monitoring device. 
The system and signal characteristics are reviewed in order to define 
the key operating features. 
The low-level algorithms are defined according to different approaches, 
starting with a simple, model free method, and progressing to 
a strategy based on the signal template and variance. 
The signal model is derived from measured data sets. 
The performance at micro-arcsec level is verified by simulation 
in conditions ranging from noiseless to large perturbations. 
\end{abstract}


\section{Introduction }
\label{sec:Introduction} 
High precision astrometry is often supported by metrology, either on 
ground or in space. 
An example of the former case is the control of atmospheric disturbances 
in long baseline interferometry, typically in the near infrared range \citep{Schmid2012,Robertson2012}, 
and of the overall instrument geometry \citep{WoillezLacour2013},
for narrow to medium angle astrometry. 
The latter case is mostly aimed at measurement or control of the 
configuration of a macroscopic instrument to a small fraction 
of wavelength, for astrometry from intermediate to large angle 
\citep{NEAT2012,SIM2008_1,SIM2008_2,GAME_EA2012}. 
The distinguishing aspect of space experiments is that usually the 
precision goal is referred to a much more compact telescope than 
ground based instruments, translating the requirements into a much 
narrower fraction of the diffraction limit; therefore, the constraints 
on image modeling \citep{GaiPASP2013} and calibration \citep{Zhai2011} 
become correspondingly more challenging.

Metrology usually employs interferometric combination of laser beams
to meet the precision goal thanks to the high photon flux. 
As opto-electronic systems become more and more complex, the 
metrology signal may be affected by a number of implementation 
aspects, potentially introducing errors. 
The data processing algorithms become then a performance factor, 
as their limitations might compromise part of the potential precision 
associated to the hardware design. 

We address a set of algorithms suited to the relative phase of 
multiple fringe patterns from two-beam laser interferometry, 
in particular the low-level, short range automated processing 
providing the relevant information to higher level diagnostics 
software and to the scientists evaluating the system response. 
The current work is focused mainly on the application to the Gaia 
metrology sub-system, although the concept is quite general and 
could be applied, in the future, to other high precision 
astrometry experiments \citep{GaiMET2012}.

The Gaia mission \citep{Perryman2005,deBruijne2012} is aimed at 
global astrometry at the few micro-arcsec (hereafter, $\mu as$) level, 
producing an all-sky catalogue of position, proper motion and 
parallax, complete to the limiting magnitude $V \simeq 20\, mag$. 
The Gaia concept relies on self-consistency of the astrometric 
information of celestial objects throughout operation, factoring 
out the instrument parameters and their evolution by calibration of 
the overall data set. 
The Hipparcos experience suggests that the approach is viable with respect 
to detection and modeling of the instrumental parameter secular evolution 
and long term variations, over time scales longer than a few revolution 
periods, i.e. above about one day. 

Gaia takes advantage of the simultaneous observations of two telescopes 
mounted on a common high stability torus to measure precisely the 
relative positions of stars in regions of the sky separated by a wide 
\textit{basic angle} $BA=106^{\circ}.5$. 
Perturbations to the instrument geometry induce displacements of each 
telescope \textit{line of sight} (LOS); the common mode displacement 
of the two telescopes does not affect the measured angular separation 
between stars, and appears as a disturbance of the overall instrument, 
with an effect similar to attitude jitter, i.e. a small degradation of 
image sharpness. 
However, the differential LOS perturbation appears as a \textit{BA 
variation} (BAV), which affects directly the astrometric measurement 
of star separation. 

The design specifications are derived from the short term astrometric 
performance requirements, corresponding to a BA stability 
$\sigma\left(BA\right)\simeq 7\,\mu as$.
In order to ensure that the short term stability requirement is actually 
met throughout operation, a custom on-board metrology system is also 
integrated in the payload design, the \textit{Basic Angle Monitoring} 
(BAM) device \citep{GaiaBAM2012_1}. 

The metrology concept is based on pairs of collimated beams from a laser 
source, split by a dedicated distribution system in front of the Gaia 
astronomical telescopes, and following an optical path very close to that 
of the stellar photons to generate an interferometric pattern on a 
dedicated section of the focal plane. 
The rationale is that perturbations to the optical configuration, affecting 
the position of the stellar images on the focal plane, i.e. the basic 
ingredient of the astrometric measurement, would affect also the BAM beam 
path, and therefore the interferogram phase. 
The match between astronomical and metrology photons is not complete, 
e.g. because the laser beam hits a much smaller region of the optical 
system, but it is assumed that the most relevant variations of the 
telescope geometry will also induce an effect on the interferogram phase, 
which is monitored, allowing for estimation of astrometric corrections 
to be applied in the data reduction, if required.

In Sec.~\ref{sec:SignalChar} we present the main characteristics
of the signal, and in Sec.~\ref{sec:AlgorDef} we define accordingly
the algorithm concepts. In Sec.~\ref{sec:Simulation} we evaluate
the performance achieved with a simple implementation of such algorithms;
then, in Sec.~\ref{sec:Discussion}, we discuss the implications
of their usage within the Gaia data reduction scheme.

\section{Signal characteristics }
\label{sec:SignalChar}
The BAM design provides a fringe pattern with period comparable with the 
stellar image size in the high resolution direction, detected by a CCD 
similar to the science ones used for object detection, astrometry 
and photometry. 
The interferogram can be considered as a large set of artificial stars, 
coherently providing an instrumental phase information. 
The BAM concept is based on two assumptions: 
\begin{itemize}
\item small amplitude of the instrument perturbations to be monitored; 
\item high signal to noise ratio (SNR). 
\end{itemize}
The latter conditions leads to a potential measurement precision corresponding
to a small fraction of the interferogram period, thus making the device
suitable to fulfill the former requirement.

The actual relationship between the interferogram phase and the telescope 
pointing direction is unknown, depending on the actual optical configuration, 
but is it assumed that, in the expected small perturbation regime, 
they will concide apart an offset, or ``zero point''. 
The BA variation, i.e. the variation of the angle between the telescope 
observing directions, is thus referred to variation of the relative phase 
between the interferograms. 
In particular, correlated phase variations are associated to common mode 
telescope pointing errors, and differential phase variations to more 
critical BA variations affecting the astrometric measurement. 

In the medium to long period, the BA variation measured by the BAM device 
can be calibrated with respect to its corresponding value on the sky, as 
deduced by convenient combination of observations, e.g. variation in the 
angular separation among stars at different epochs. 
The comparison between BAM measurements and astrometry will 
therefore provide an assessment of science quality throughout 
the mission.

The interferogram associated to an ideal BAM, feeding an ideal 
telescope, is the point spread function of the unobstructed circular 
aperture related to the individual laser beams, modulated by fringes 
with the Young period. 
Practical implementation introduces deviations from such simple 
framework, because of optical aberrations due to design, manufacturing 
and alignment in the metrology distribution system and in the telescope. 
Also, the laser beam intensity is not uniform over its cross section, 
but it is expected to be described e.g. by a truncated Gaussian. 

This still generates a useful signal \citep{GaiaBAM2012_2} 
with high contrast fringes within a more complex diffraction 
envelope. 
The fringe period is basically preserved, but the phase modulation 
bears contributions from the real system. 
Also, the envelope has generally larger size, and significant shape 
variation, with respect to the ideal case. 
However, it is expected that the crucial information of BA variation is 
still encoded in the mutual relationship between interferograms. 
In particular, the small BA variations expected during operation will 
introduce interferogram displacements on the $\mu as$ scale, with amplitude 
much smaller than the fringe period. 

The BAM device, by design, generates an interferogram for each telescope,
imaged in different positions of a common dedicated CCD. Each signal
is affected by its own photon shot noise, background (depending on the 
current pointing position), and comparable readout noise. 
Photon noise is expected to be the main contribution, since the laser 
source intensity is tuned in order to use most of the detector dynamic 
range. 

\begin{figure}
\includegraphics[width=75mm,height=50mm]{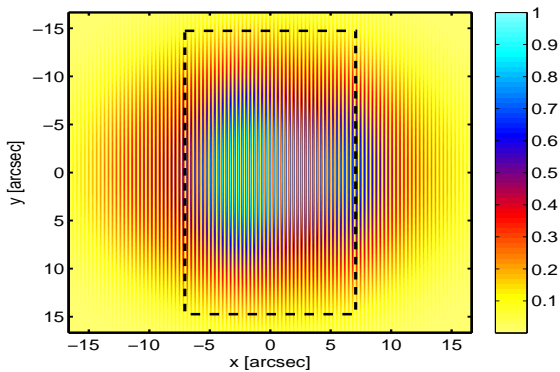} 
\caption{\label{fig:FringePattern2D}Gaia BAM interferogram; dashed 
rectangle: readout area }
\end{figure}

\begin{figure}
\includegraphics[width=75mm,height=50mm]{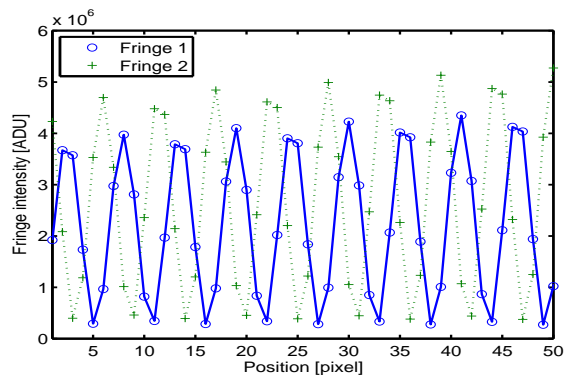} 
\caption{\label{fig:FringePattern1D}Section of two one-dimensional 
fringe patterns, integrated across fringe }
\end{figure}

The BAM is based on a geometry similar to that of Young's interference 
experiment, in which the two slits are replaced by small beams with 
diameter $D=9\, mm$, separated by a baseline $B=0.54\, m$, generating 
on the focal plane a modulated signal, i.e. an interferogram, with 
$\sim 2.44 B/D=146$ fringes, limited by the diffraction envelope of 
the individual aperture. 
The two telescopes are fed by separate beam pairs, generating similar 
interferometric spots on the detector; the fringe period is 
$\lambda/B=325\, mas$, corresponding to $55\,\mu m$, i.e. $5.5$~pixels.
The interferogram from each telescope is repeatedly measured,
at a rate of an image every $\sim 20\, s$. 

Hereafter, we will refer to the beam pair monitoring one telescope,
to the related optical system, and to the corresponding detected signal,
as a ``BAM channel''. Only a limited region of interest over the
central lobe of diffraction is actually acquired, for each BAM channel,
as shown by the dashed rectangle in Fig.~\ref{fig:FringePattern2D}.
The interferogram is integrated across the fringes after readout,
in order to achieve a more easily tractable one-dimensional signal,
shown in Fig.~\ref{fig:FringePattern1D}. 
Since the fringe patterns are mostly aligned to the CCD array, to 
within $1^{\circ}.5$, only a marginal smearing is introduced.

\section{Algorithm definition }
\label{sec:AlgorDef}
The laser source is monochromatic, but the interferometric
signal, due to the limited beam diameter and associated fringe envelope
size, has a spatial frequency distribution spread over the approximate
range $\left[\frac{\lambda}{B+D},\frac{\lambda}{B-D}\right]$. 
A detailed mathematical description of the fringe pattern is not 
easily formulated except in the simplest geometric cases, due to 
optical aberrations.

However, most phase estimation algorithms are based either on a 
template, or on assumptions on the mathematical characteristics of 
the signal, i.e. they are \textit{model dependent}. 
Given the BAM readout frequency and high SNR, it is possible 
to define a numerical \textit{signal template}, as well as its 
empirical variance, both deduced from a convenient set of measured 
data, rather than analytical formulation of a signal model whose 
parameters have to be estimated e.g. by best fit to the measurements.

It is possible to define algorithms aimed at providing a phase 
estimate for each interferogram, related to the average LOS position 
of the individual telescope, from which the BAV is simply deduced by 
difference of the phase values at each time. 
We focus on a set of simple concepts leading to convenient algorithm 
implementation, and evaluate their main impact on the measurements. 

We also address a \textit{model independent} measurement concept, 
providing a direct estimate of the BAV as phase difference between 
the fringe patterns, without individual phase computation. 

A model dependent approach is expected to be, in principle, more precise,
because the expected characteristics of the signal are taken into
account. However, this requires that additional information is available
to define at least the signal template and variance, as in Sec.~\ref{sub:TemplateDef}.
Hereafter we expand on some of the possible approaches,
which will then be tested on simulated data.

\subsection{Template definition from the data }

\label{sub:TemplateDef}
From a sufficient number $N$ of measurement
of the interferometric signals $S_{1},\, S_{2}$ related to each telescope,
it is possible to define the sample approximations $T_{1},\, T_{2}$
of the corresponding templates, or noise free reference functions,
from the signal averaged over the data set: 
\begin{equation}
T_{1,2}\left(x_{k}\right)=\frac{1}{N}\sum_{n=1}^{N}S_{1,2}\left(x_{k};t_{n}\right)\,.\label{eq:Template1}
\end{equation}
The dependence on the pixel position $x_{k}$ and on the current exposure
time $t_{n}$ has been shown explicitly, but it will be omitted below where
not necessary for the sake of simplicity. 

We define in a similar way the empirical variance
$V_{1},\, V_{2}$ of each interferogram pixel as 
\begin{equation}
V_{1,2}\left(x_{k}\right) = 
\frac{1}{N-1}\sum_{n=1}^{N}\left[S_{1,2}-T_{1,2}\right]^{2} 
\, . 
\label{eq:SigVar}
\end{equation}

This approach is based on the assumption that the measurement conditions
are stationary, so that the average can be considered as a reasonable
approximation of the template. 

The phase noise on the fringes is assumed to be much smaller than
the fringe period, by a factor of $\sim10^{-5}$, in normal conditions.
However, phase noise will induce errors on the template and variance 
estimate, in addition to the amplitude fluctuations due to photon noise. 
It can be shown that photon noise, in the expected conditions, is 
dominant; thus, the small fringe displacement among subsequent instances 
does not introduce a significant blurring of the mean fringe, which can 
be considered a reasonable approximation to the desired signal template. 

In the numerical experiments below, the template is built by averaging
over $1,000$ frames, so that the expected reduction of individual
measurement noise is of order of $\sqrt{1000}\simeq31$. 

The data set also provides an estimate of the signal dispersion through
the sample variance, so that subsequent measurements can then be checked
for consistency with the template and the initial data (e.g. in terms
of confidence levels). 
For each BAM channel, a specific template is defined by averaging 
the corresponding data set from a suitable measurement sequence.

\subsection{Model free fringe location: Mutual Correlation (MC) }
\label{sub:MutualCorr}
The signals from the two BAM channels are not exactly sinusoidal,
but they have a very narrow spectrum, resulting in a significant modulation
at the same nominal spatial frequency. Their envelopes are not equally
shaped, but over the readout region their amplitude is quite high.

The signal similarity leads to the possibility of processing them
with a simple approach based on the correlation technique. The two
quasi sinusoidal signals will have a maximum correlation value for
a given displacement with respect to each other, which in the example
shown in Fig.~\ref{fig:FringePattern1D} is of order of two pixels.
A perturbation to whichever telescope, inducing a phase variation
on the corresponding interferogram, will modify accordingly the position
of the signal correlation maximum. Therefore, this algorithm defines
a linear variable strictly related to the relative position of the
interferograms, and as a consequence of the basic angle variation between
telescope lines of sight.

\textit{We adopt the estimate of this relative interferogram phase
as an operational definition of the BA estimate, in order to monitor
its variation around an irrelevant zero point to be calibrated on
the sky from science data. } 
This approach, by construction, gives a position estimate 
modulo the fringe period, which is not an ``absolute'' datum.

The relative displacement of the two fringe patterns can be estimated
directly by computation of the maximum correlation position between
them, e.g. from parabolic fit of the three correlation values achieved
for lag zero and $\pm 1$~pixel. This approach completely avoids
the issue of template definition; broadly speaking, the interferogram
from each channel acts as a template for the other. In practice, the
method does not need that the two signals are very similar to each
other (which would improve on the correlation value), but only that
their shape remains approximately stable over the time frame required
for calibration on the sky through the science data.

In more detail, the two signals are arrays of 500 values; given 
their offset of about two pixels, shown in Fig.~\ref{fig:FringePattern1D}, 
they are roughly ``aligned'' by removing the first value on the former 
and the last one on the latter, i.e. reducing the data to $K = 499$ 
values. 
The offset is due to the simulated placement of readout windows; it is 
constant over the simulation, due to the small phase noise. 

The lag zero signal is then selected as the central part of the array 
with index $k = 2,\ldots,K-1$, i.e. excluding the end values. 
The lag $\pm 1$ signals include respectively either of the end values, 
discarding the opposite ending value, i.e. with indices 
$k = 1,\ldots,K-2$ and $k = 3,\ldots,K$. 
The correlation is then computed over $K-1 = 498$ values. 
\\ 
This windowing approach sacrifices a small part of the signal, but 
it avoids data interpolation, or extrapolation at each end 
of the arrays. 
Processing is fast, as only array indexing is used. 

The mismatch between signals from each BAM channel suggests that the
noise performance of this algorithm may not be optimal, but its advantages
are the simplicity and robustness, so that it is applicable also in
a context of limited knowledge of the instrument parameters, e.g.
at the beginning of operation or after unexpected events inducing
large perturbations. It also provides a simple sanity check for other
algorithms.

\subsection{Model dependent fringe location: Correlation with Template (CT)}
\label{sub:CorrTempl}
In the expected operating conditions, the interferometric signal is
quite stable, therefore we assume that it is possible to apply the
approach of Sec.~\ref{sub:TemplateDef} to define its model (or template)
by averaging the measurements according to Eq.~\ref{eq:Template1}. 
As far as only small perturbations are present, the individual signal 
instances will be consistent with such signal template, apart from 
fluctuations in amplitude (e.g. from photon noise) and phase (related 
to the BA variation).

For any fringe pattern instance, we compute the correlation with the
template, in its nominal position, and with $\pm 1$~pixel offset. 
The indexing of Sec.~\ref{sub:MutualCorr} is used for definition 
of the lag zero and $\pm 1$ signals. 
Also, parabolic fit of the correlation values provides the 
``exact'' location of the current interferogram with respect to 
its template. 
\textit{We adopt this parameter as an estimate of the relative LOS 
position, independently for each BAM channel. } 
The difference between LOS values is considered the operational definition 
of the BA, so that its variation can be used for monitoring.

In stable conditions, each LOS is expected to fluctuate around zero,
since each signal instance is close to its template within the noise;
similarly, the BA estimate is expected to be centered in zero.

\subsection{Model dependent fringe location: Maximum Likelihood estimator (ML) }
\label{sub:MaxLik} 
Correlation is a robust and proven technique, but it is known to provide
sometimes less than optimal noise performance. We define an estimator
derived in the maximum likelihood framework, which also requires knowledge
of the current noise statistics, in particular the signal variance,
estimated on the data as from Eq.~\ref{eq:SigVar}.

Given two similar functions, hereafter labeled ``signal'' $S$ and
``template'' $T$, the maximum likelyhood approach may be used to
identify the best matching position, minimizing a functional inspired
to the classical $\chi^{2}$ and defined as 
\begin{equation}
\chi^{2} = 
\sum_{k=1}^{K}\frac{\left[S_{k}-T\left(x_{k}-\tau\right)\right]^{2}}
{\sigma_{k}^{2}}
\, , \label{eq:chi2}
\end{equation}
 where the dependence on the coordinates has been shown explicitly only
for the template. The matching position has true value $\tau_{0}$,
and its estimate $\tau$ is derived from the measured data. The derivation
follows that in \citet{GaiPASP1998}, 
in which the signal $S_{k}=S\left(x_{k}\right)$, measured over the
$K$ pixel positions $x_{k}$, was considered as equal to the template
$T$ apart an unbiased measurement error and a shift in the reference
position, therein labelled ``photo-centre''. It may be noted that,
in case of usage of different functions, e.g. a template not matching
the signal, or an evolving signal shape (non stationary conditions),
the signal discrepancy 
\begin{equation}
h_{k} = S_{k}-T_{k} \, 
\label{eq:discrepancy}
\end{equation}
 is no longer just due to noise, so that it cannot be expected in
general to be uncorrelated and to have zero mean.

The photo-centre estimate must be a stationary point of the $\chi^{2}$
in Eq.~\ref{eq:chi2}, i.e. a solution of the equation 
\begin{equation}
\sum_{k}\frac{\left[S_{k}-T_{k}\left(\tau_{0}\right)\right]\cdot T_{k}'\left(\tau_{0}\right)}{\sigma_{k}^{2}\left(\tau_{0}\right)} 
= 0 \, . 
\label{eq:Solution1}
\end{equation}
 Therefore, assuming small errors ($\tau\approx\tau_{0}$), the square
bracket can be simplified, taking advantage of Taylor's expansion
of the template in the current approximate position, to provide 
\begin{equation}
\tau-\tau_{0} = 
-\frac{\sum_{k}\frac{\left[S_{k}-T_{k}\left(\tau\right)\right]\cdot 
T_{k}'\left(\tau\right)}{\sigma_{k}^{2}\left(\tau\right)}} 
{\sum_{k}\frac{\left[T_{k}'\left(\tau\right)\right]^{2}}
{\sigma_{k}^{2}\left(\tau\right)}} \, . 
\label{eq:Solution2}
\end{equation}
 The solution is unbiased, i.e. $\left\langle \tau-\tau_{0}\right\rangle 
\equiv 0$, and has variance equal to 
\begin{equation}
\left\langle \left(\tau-\tau_{0}\right)^{2}\right\rangle = 
\left[\sum_{k}\frac{\left[T_{k}'\left(\tau\right)\right]^{2}}
{\sigma_{k}^{2}\left(\tau\right)}\right]^{-1} \, . 
\label{eq:SolVariance}
\end{equation}
It may be noted that at increasing values of location error, due 
e.g. to lower SNR, the validity of the above expressions 
degrades progressively. 
\textit{The above Eq.~\ref{eq:Solution2}
is adopted as operational definition of the ML estimate of LOS, for
each BAM channel, and the BA is defined as the LOS difference. }

\subsection{Residual diagnostics }
\label{sub:Chi2Red}
In order to assess the consistency of individual
interferogram $S_{k}$ with the current estimate of signal template
$T$ and of its variance $V$, from Eqs.~\ref{eq:Template1} and
\ref{eq:SigVar}, we define a quantity related to the reduced $\chi^{2}$
concept according to Eq.~\ref{eq:chi2}: 
\begin{equation}
\chi_{R}^{2}=\frac{1}{K-1}\,\sum_{k=1}^{K}\frac{\left[S_{k}-T\left(x_{k}-\tau\right)\right]^{2}}{\sigma_{k}^{2}}\,.\label{eq:chi2red}
\end{equation}
 When all residuals are comparable with the expected noise, corresponding
to the signal variance, the reduced $\chi^{2}$ is expected to be
of order of unity. The normalisation factor $K-1$ is used under the
assumption that no model parameter is derived from the current measurement,
apart an overall photometric level estimate. The issue of defining
the actual number of degrees of freedom is quite sensitive, in general,
and may require some care in practical implementation.

However, we may expect that the value of $\chi_{R}^{2}$ as defined
in Eq.~\ref{eq:chi2red} might at least help in identifying significant
perturbations, when it becomes significantly larger than unity.

\section{Simulation }
\label{sec:Simulation}
The algorithm performance is evaluated, first of all, in terms of noise 
sensitivity with respect to signal amplitude, and of linearity with respect 
to signal phase; then, on externally generated data, with respect to random 
noise and systematic error performance. 
The processing is performed on a desktop PC, endowed with an Intel Xeon 
3.33~GHz CPU and 8~GB RAM, using the Windows~7 (64~bit) operating system, 
and the Matlab package.

\subsection{Performance as a function of amplitude noise }

\label{sub:NoisePerf}The phase noise performance is simulated for
the three algorithms defined above, using signal profiles considered
as realistic for an optical configuration affected by a plausible
amount of manufacturing, mounting and alignment errors.

\begin{figure}
\includegraphics[width=75mm,height=55mm]{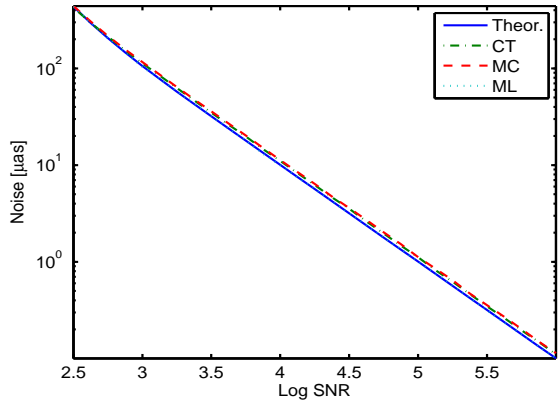} 
\caption{\label{fig:NoisePerf1}Algorithm precision as a function of SNR }
\end{figure}

Over a range of intensity corresponding to increasing SNR, from few
hundreds to about one million, we estimate the theoretical location
precision for the ML algorithm according to Eq.~\ref{eq:SolVariance}.
For each algorithm, we generate a noisy signal sample of $N=10,000$
instances, corresponding to constant interferogram phase and random
amplitude fluctuations related to the photon noise level associated
to the current intensity level. The fringe position, or the corresponding
interferogram separation, is estimated according to a simple implementation
of each algorithm (MC, CT, ML). The corresponding BA noise due to
amplitude fluctuations is then evaluated as the RMS dispersion over
the sample. The result is shown in Fig.~\ref{fig:NoisePerf1}.
Over most of the SNR range, the trend is basically photon limited,
as evidenced by the constant slope in logarithmic units. 

\begin{figure}
\includegraphics[width=75mm,height=55mm]{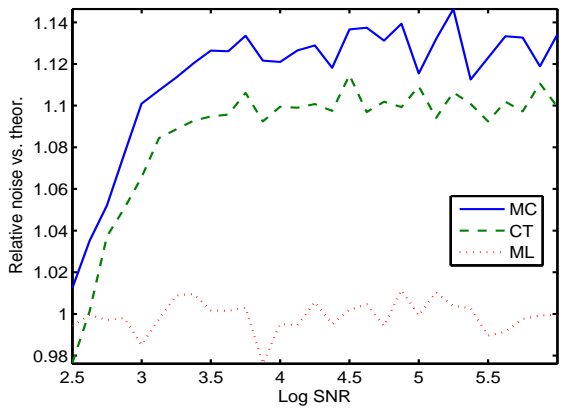} 
\caption{\label{fig:NoisePerf2}Noise with respect to the ML case }
\end{figure}

The noise decreases with increasing SNR, as expected, and the algorithm
performance difference cannot be easily distinguished on this scale.
In order to ease the comparison of algorithm results, the relative
precision is shown, referred to the estimated precision from Eq.~\ref{eq:SolVariance},
which can be considered as a conservative estimate of the limiting
performance, providing a good match with the Maximum Likelihood location
estimator defined in the same framework. The result is shown in Fig.~\ref{fig:NoisePerf2}.

In the high SNR regime, the noise performance of both correlation
methods is comparable, and about $10\%$ (CT) and $12.5\%$ (MC) worse
than the limiting error defined in the Maximum Likelihood framework.
The fluctuations of the relative performance are compatible with the
statistics related to the sample size, i.e. of order of $1\%$.

\subsection{Linearity performance }
\label{sub:LinPerf}
\begin{figure}
\includegraphics[width=75mm,height=55mm]{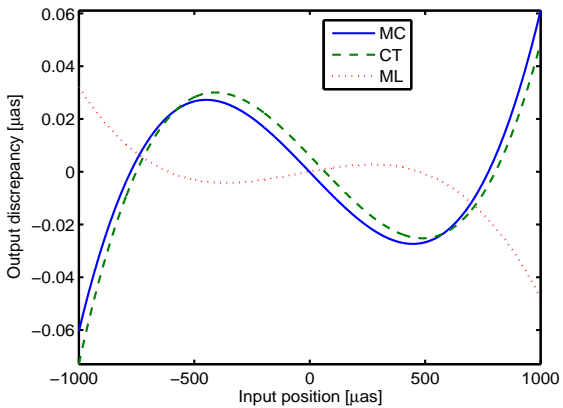} 
\caption{\label{fig:Linearity}Input/output discrepancy for ML, CT and MC }
\end{figure}

The algorithm linearity is evaluated in a noiseless
case by simulation, injecting a known LOS displacement on the input
signals and verifying the LOS estimate in output (for CT and ML).
For the MC algorithm, the BAV estimate is averaged for input displacements
applied to either LOS1 or LOS2 (both giving quite similar, but not
exactly equal, results). The test range is $\pm 1\, mas$, i.e. quite
large with respect to the small perturbations expected in the normal
operating regime. 
A linearity correction was required for both CT and MC. 
The input/output discrepancy of ML, and the
corresponding residual discrepancy after linearity correction for
CT and MC, are shown in Fig.~\ref{fig:Linearity}.

The linearity of all algorithms, after correction in the case of MC
and CT, is quite good, providing a residual discrepancy of a few 
$0.01\,\mu as$ over the $\pm 1000\,\mu as$ range considered.

\subsection{Performance of the algorithms on externally simulated data }

Two data sets including a number of realistic effects have been generated
by the Coordination Unit 2 (CU2) of the Gaia Data Processing and Analysis
Consortium (DPAC), on the supercomputer Mare Nostrum at the Barcelona 
Supercomputing Center (Centro Nacional de Supercomputaci\'on), including 
contributions from different real world sources. 
They are processed according to the algorithms in Sec.~\ref{sec:AlgorDef}. 
The data sets include respectively 5,181 and 3,638 interferogram pairs
representing sequences of BAM exposures.

\subsubsection{Overall features of the data set 1 }

\label{sub:OverallDS1}The signal to noise ratio of this set of fringe
pattern pairs is extremely high, of order of $6 \times 10^{5}$ and
$10^{6}$ respectively for the whole LOS 1 and LOS 2 signals. From
Sec.~\ref{sub:NoisePerf}, this corresponds to an astrometric noise
level of order of $0.2\,\mu as$ and $0.1\,\mu as$, respectively.

\begin{figure}
\includegraphics[width=75mm,height=55mm]{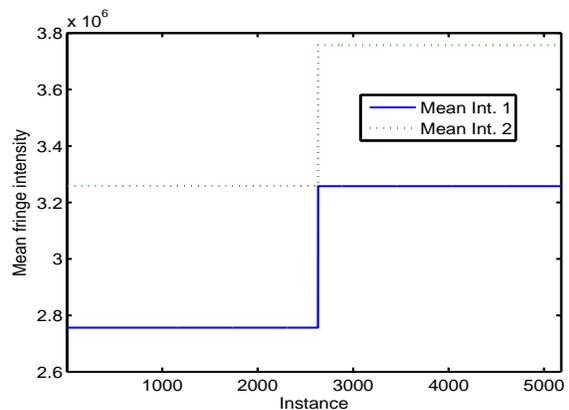} 
\caption{\label{fig:IntDiscont1}Mean fringe intensity, data set 1}
\end{figure}

At an early analysis, a discontinuity in the data set was evidenced,
as can be seen in Fig.~\ref{fig:IntDiscont1}, showing the average
intensity over each interferogram (i.e. the mean level of the fringe
pattern). This intensity variation might correspond e.g. to a variation
of the laser source intensity, although the amplitude of the variation
($\sim 20\%$) and its suddenness may not be expected as a realistic
common event. However, it is interesting to check the robustness of
our diagnostics and measurement algorithms against such variations.

\begin{figure}
\includegraphics[width=75mm,height=55mm]{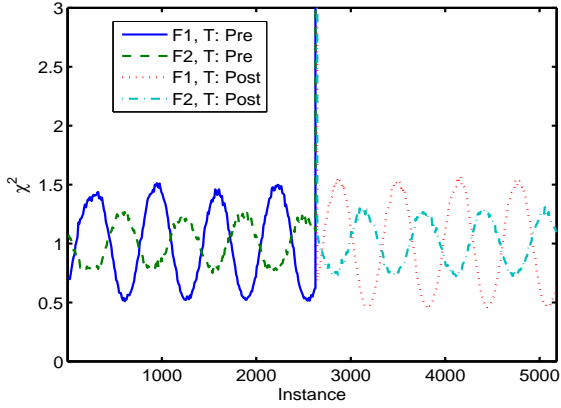} 
\caption{\label{fig:chi2_DS1_2}Reduced $\chi^{2}$, zoom on unity; 
data set 1 }
\end{figure}

Given the relative stability of the signal level on each side of the
discontinuity, we compute the signal template and variance separately
for each subset, according to Eqs.~\ref{eq:Template1} and \ref{eq:SigVar};
they are labeled respectively ``Pre'' and ``Post'', with reference
to the discontinuity. According to Sec.~\ref{sub:Chi2Red}, we evaluate
the reduced $\chi^{2}$ for each interferogram using both templates.
The results are shown in 
Fig.~\ref{fig:chi2_DS1_2}.

We remark that the large signal amplitude variation in 
Fig.~\ref{fig:IntDiscont1}
is evidenced also by a huge $\chi^{2}$ variation, by 
about seven orders of magnitude, when processing the data 
on either side of the discontinuity with the mismatched template. 
The $\chi^{2}$ increase is consistent with the intensity variation 
and the SNR level: the signal change is order of $10^{4}$ 
times larger than its variance. 
Besides, using matched data and templates, i.e. ``Pre'' template with 
``Pre'' data and ``Post'' template with ``Post'' data, the reduced 
$\chi^{2}$ value remains of order of unity, as shown in 
Fig.~\ref{fig:chi2_DS1_2}.

\subsubsection{LOS and BAV of the data set 1 }

\begin{figure}
\includegraphics[width=75mm,height=55mm]{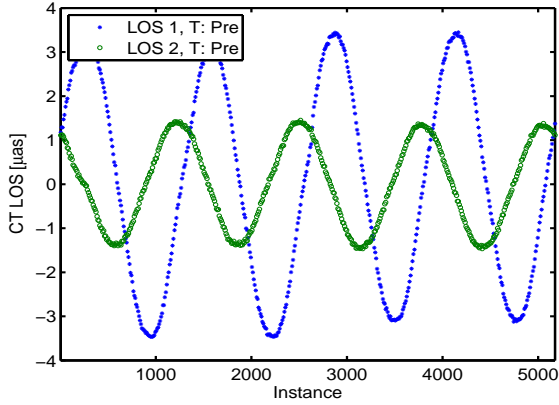} \caption{\label{fig:LOS_CT_DS1}LOS estimate by CT; data set 1}
\end{figure}

\begin{figure}
\includegraphics[width=75mm,height=55mm]{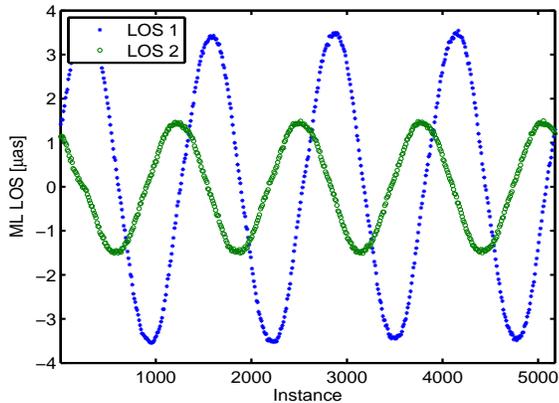} \caption{\label{fig:LOS_ML_DS1}LOS estimate by ML; data set 1}
\end{figure}

The LOS evolution over the data set 1 are shown in Fig.~\ref{fig:LOS_CT_DS1},
estimated using the CT algorithm. The linearity correction from Sec.~\ref{sub:LinPerf}
is applied. In Fig.~\ref{fig:LOS_ML_DS1} the LOS estimate from the
ML algorithm is shown. A crude background subtraction is applied,
by removing the fringe mean value. The zero points have been set according
to display convenience; also, each display point replaces 10 initial points with 
their average, also reducing the noise, in order to improve on plot readability.

\begin{figure}
\includegraphics[width=75mm,height=55mm]{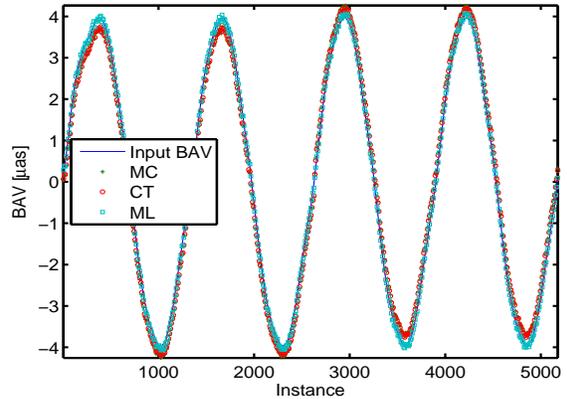} \caption{\label{fig:BAV_MC_CT_ML}Input and estimated BAV; data set 1}
\end{figure}

The ML result is quite similar to that from CT for both LOS~1 and
LOS~2. The two algorithms appear therefore to be quite consistent
with each other. The signal discontinuity appears to induce a marginal
effect on both LOS~1 and LOS~2, corresponding to an offset of a
few $0.1\,\mu as$. 
Both LOSs feature an approximately sinusoidal behaviour, with different 
phase, and period corresponding to the satellite revolution (about six 
hours). 
The oscillations are referred to simulation of the thermo-elastic 
evolution of the instrument during the spin, i.e. a physical phenomenon; 
the two telescopes are mounted in different positions and therefore 
respond independently to a given external perturbation (e.g. residual 
effects of Sun irradiation). 
Besides, the LOS and BAV ``jump'' is due to the change in the signal 
profile between either side of the discontinuity, filtered by the 
truncation due to readout windowing and the algorithm response.

The sequence of BAV estimates from each algorithm is shown in Fig.~\ref{fig:BAV_MC_CT_ML},
also including the nominal BAV values used in input to the simulation
(solid line). It may be noted that all algorithms reproduce a large
part of the input BA oscillations, to a few $0.1\,\mu as$.

\begin{figure}
\includegraphics[width=75mm,height=55mm]{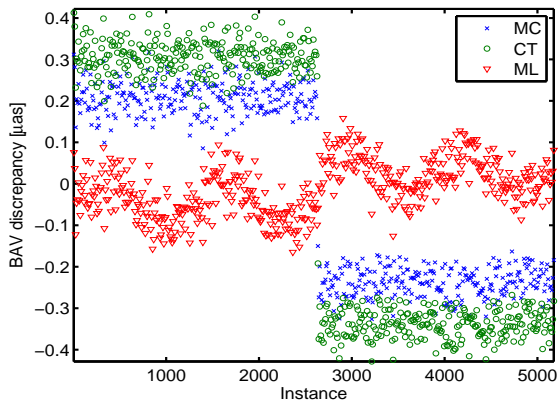} 
\caption{\label{fig:dBAV_MC_CT_ML}Input/output discrepancies; 
data set 1}
\end{figure}

The BAV discrepancy from the input value is shown in Fig.~\ref{fig:dBAV_MC_CT_ML}
for each algorithm. 
The effect of signal discontinuity is a small BAV ``jump'' (by a 
few $0.1\,\mu as$) on CT and MC estimates (circles and crosses, 
respectively), and hardly perceivable ($\sim 0.1\,\mu as$) on the 
ML results (triangles). 

The mean and RMS discrepancy of the results from each algorithm with respect 
to the ``true'' input BAV, and of the result discrepancy between algorithms, 
are listed in Table~\ref{tab:AlgDiscr1} separately for the two region.
The input/output BAV discrepancy is the algorithm error (a few $0.1\,\mu as$ 
for MC and CT, a few $0.01\,\mu as$ for ML), whereas the output difference 
(a few $0.1\,\mu as$) is related to the mutual algorithm consistency. 
The RMS values can be considered as the random noise of the measurement, 
whereas the offsets represent a systematic difference. 

The noise is of order of $0.1\,\mu as$, as expected, within each
data region (``Pre'' or ``Post''); the average value is quite
close to zero for ML, evidencing a very small systematic error, 
consistent with the very small ``jump'' 
(Fig.~\ref{fig:dBAV_MC_CT_ML}). 
The results from the different algorithms are consistent to within 
a few $0.1\,\mu as$. 

The MC and CT results are grouped along straight lines 
separated by a vertical offset, corresponding to a ``jump'' 
(associated to the signal discontinuity) of $\sim 0.5\,\mu as$, 
significantly larger than the intrinsic dispersion in either 
``Pre'' or ``Post'' region. 

\begin{figure}
\includegraphics[width=75mm,height=55mm]{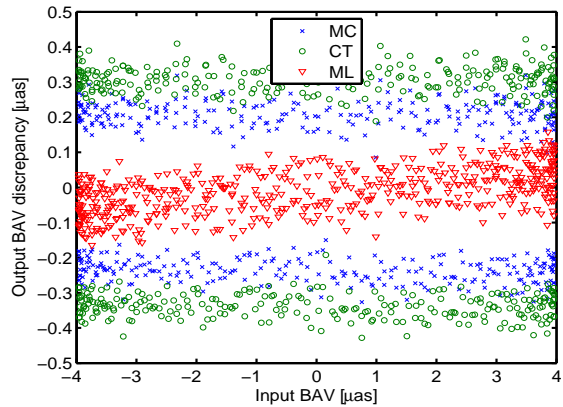} \caption{\label{fig:BAV_DS1_lin}BAV discrepancy vs. input BAV }
\end{figure}

The algorithm linearity can be verified by direct comparison of input 
and output values; the ideal case of perfect matching between input and 
output would correspond to zero discrepancy. 
The BAV discrepancy is shown in Fig.~\ref{fig:BAV_DS1_lin}
for all three algorithms. 
Their response is quite linear, since the slope of the output discrepancy 
vs. input BAV is very small; the offsets are again due to the signal 
discontinuity.
The analysis of linear correlation between input and output provides 
a correlation coefficient in all cases above $99.9\%$, and a slope 
very close to unity, for all algorithms. 

\subsubsection { Results on data set 2 } 
The SNR of this data set is still very high with respect to usual 
astronomical data, but significantly lower than the previous case, 
of order of $5 \times 10^{4}$ for both LOS~1 and LOS~2 signals, 
corresponding to the nominal BAM operation level. 
From the results of Sec.~\ref{sub:NoisePerf}, this corresponds to 
an expected astrometric noise level of order of $1\,\mu as$. 

The simulation results evidence a noise level of $\sim 1.4\,\mu as$, 
consistent with the expectations. 
The algorithm behaviour with respect to signal perturbations, including 
intensity increase by $14\%$ for channel 1, and $12\%$ for channel 2, is 
again limited to a BAV ``jump'' of about $1\,\mu as$. 
The results from MC, CT and ML remain quite consistent with each other. 
Also, the signal variation is clearly evidenced by the reduced 
$\chi^{2}$.

\section{Discussion }
\label{sec:Discussion}
The performance of low-level algorithm on BAV noise and linearity has 
been verified on different data sets at the $\mu as$ level or better, 
consistently with the SNR limit.
In spite of its simplicity, the model free approach (MC) achieves 
noise performance within a few percent of the template based CT
and within $\sim 12.5\%$ of the ML limit. 
MC results are also quite consistent with those from CT, but with 
lower sensitivity to measurement ``jumps'' induced by some signal 
changes. 
The ML approach takes full advantage of the available information,
thus resulting in best noise performance and very little sensitivity
to the perturbations simulated in the data sets used. 

The availability of different algorithms with comparable sensitivity, 
and somewhat different response to signal disturbances, may help 
in detecting perturbations throughout operation by evidencing their 
effects, and minimising their impact on the science measurements by 
appropriate corrections introduced in the data reduction. 
The low-level algorithms do not perform any interpretation of the signal 
variation simulated in data set 1, in terms of estimation of relevant 
system parameter change, however they provide a clear diagnostics of 
its occurrence through the $\chi^2$, and the BAV estimate 
error is retained within the specifications. 

The reduced $\chi^{2}$ evidences oscillations around unity value 
on the data set 1 (Fig.~\ref{fig:chi2_DS1_2}), even using the 
matching template and variance for each data subset. 
This is assumed to be related to the extremely high SNR, and 
correspondingly very low measurement noise. 
In such circumstances, the simple procedure used to build the signal 
template and variance seems to be insufficient, e.g. because the input 
phase variation induces signal variations not negligible with respect 
to those associated to photon noise. 
The fluctuations on ML estimates (Fig.~\ref{fig:dBAV_MC_CT_ML}), 
in phase with the BAV, appear to be consistent with this scenario. 
Such oscillations are no longer present in the reduced $\chi^{2}$ 
and BAV computed on data set 2. 

The current simple approach appears to be adequate to the Gaia 
mission requirements, even in spite of the large simulated signal 
variations, resulting in systematic errors compatible with the 
micro-arcsec noise level associated with the SNR range expected 
for the Gaia BAM operation. 
It may be noted that, with the Gaia pixel size $\sim 60\, mas$,
the best precision case of data set 1, of order of $0.1\,\mu as$, 
corresponds to a few micro-pixels, thus matching well the 
simulations and experimental results in \citet{Zhai2011}.

The minimisation of signal model error, according to 
Sec.~\ref{sub:TemplateDef}, requires large amounts of data to 
average out the noise fluctuations. 
Besides, this may reduce the sensitivity to disturbances acting on 
a time scale shorter than the period of data accumulation. 
A suitable trade-off must therefore be defined.

\section{Conclusions }
\label{sec:Conclusions} 
We investigate algorithms aimed at low-level processing of metrology 
signals from two laser beam, high dilution, imaging metrology systems, 
with particular reference to the Gaia BAM device. 
The methods range from a model free approach based on mutual correlation 
(MC), to others using progressively more detailed information, including 
the signal template (CT), and the signal variance (ML). 

The numerical model of the signal is derived by computing the average
fringe pattern and its variance over a convenient data set, which
may be easily verified for self-consistency, and has minimal dependence
on external parameters: mainly pixel size and optical scale to convert
linear results (fractions of pixel) into angular values.

The performance is verified at or below the $\mu as$ level by simulation 
on data sets generated either in house, or by independent groups, ranging 
from the noiseless case to comparably large perturbations. 
The model free approach still provides quite appealing performance, while 
the maximum likelihood method features lowest noise and sensitivity 
to disturbances. 
Some signal disturbances, including profile variation and up to $20\%$ 
intensity variation, induce discontinuities in the measured positions of 
order of $1\,\mu as$, within the design specifications. 
The signal variation is anyway clearly identified by $\chi^2$ diagnostics. 

The performance of all low-level algorithms appears therefore to be of 
interest for application to the data reduction of the Gaia Basic Angle 
Monitoring device, and potentially to future high precision astrometry 
experiments. 

\acknowledgments 
The activity described in this paper is partially supported by the 
contract ASI I/058/10/0. 
We acknowledge the contribution of the Coordination Unit 2 (CU2) of 
the Gaia Data Processing and Analysis Consortium (DPAC) for the 
generation of the BAM data used for some of our tests. 
Such data have been simulated on the supercomputer Mare Nostrum at 
Barcelona Supercomputing Center (Centro Nacional de 
Supercomputaci\'on).

\clearpage

\begin{table}
\begin{center}
\caption{\label{tab:AlgDiscr1}Algorithm accuracy and consistency 
before and after the signal discontinuity; data set 1} 
\begin{tabular}{lcccccc}
\tableline
\tableline
 & CT error & MC error & ML error & CT - MC  & CT - ML  & MC - ML \\
 & [$\mu as$] & [$\mu as$] & [$\mu as$] & [$\mu as$]  & [$\mu as$]  & [$\mu as$] \\
\tableline
Mean Pre & -0.231  & -0.272  & -0.046  & 0.041  & -0.185  & -0.227 \\
RMS Pre  & 0.127  & 0.127  & 0.138  & 0.012  & 0.059  & 0.055 \\
Mean Post & 0.218  & 0.260  & 0.026  & -0.043  & 0.191  & 0.234 \\
RMS Post  & 0.107  & 0.106  & 0.118  & 0.012  & 0.059  & 0.055 \\
\tableline
\end{tabular}
\end{center}
\end{table}

\end{document}